\documentclass[fleqn,usenatbib]{mnras}

\usepackage{newtxtext,newtxmath}
\usepackage[T1]{fontenc}
\usepackage{ae,aecompl}
\usepackage{graphicx}
\usepackage{amsmath}
\usepackage{amssymb}

\newcommand{\kepler}{\textit{Kepler}}
\newcommand{\Gaia}{\textit{Gaia}}
\newcommand{\ktwo}{\textit{K2}}

\newcommand{\thalf}{$t_{1/2}$}

\newcommand{\gaia}{\textit{Gaia}}

\newcommand{\orionage}{$4.0\pm0.2$}
\newcommand{\minflareenergy}{$6.7\times10^{33}$}
\newcommand{\maxflareenergy}{$5.2\times10^{35}$}
\newcommand{\minmaxflareenergy}{$2.5\times10^{34}$}
\newcommand{\maxxrayflareenergy}{$7.7\times10^{34}$}
\newcommand{\minmaxxrayflareenergy}{$1.4\times10^{33}$}
\newcommand{\flarenumber}{26}

\newcommand{\Msun}{$M_{\odot}$}
\newcommand{\kdwarfnumber}{10}
\newcommand{\mdwarfnumber}{49}
\newcommand{\energylimit}{\rewrite{$1\times10^{35}$}}
\newcommand{\rewrite}[1]{\textcolor{black}{#1}}
\newcommand{\rewritetwo}[1]{\textcolor{black}{#1}}
\newcommand{\rewritethree}[1]{\textcolor{black}{#1}}

\title[White-light flares in Orion]{NGTS clusters survey - II. White-light flares from the youngest stars in Orion}

\author[J. A. G. Jackman et al.]{James A. G. Jackman,$^{1,2}$\thanks{E-mail: J.Jackman@warwick.ac.uk}
Peter J. Wheatley,$^{1,2}$\thanks{E-mail: P.J.Wheatley@warwick.ac.uk}
Jack~S. Acton,$^{3}$\newauthor
David R. Anderson,$^{1,2}$
Claudia Belardi,$^{3}$
Matthew~R. Burleigh,$^{3}$\newauthor
Sarah~L. Casewell,$^{3}$
Philipp Eigm\"uller, $^4$
Samuel Gill,$^{1,2}$
Edward Gillen,$^{5,\ddagger}$\newauthor
Michael~R. Goad,$^{3}$
Andrew~Grange,$^{3}$
Simon~T. Hodgkin,$^{6}$
James~S. Jenkins,$^{7,8}$\newauthor
James McCormac,$^{1,2}$
Maximiliano Moyano,$^{10}$
Didier  Queloz,$^{5}$
Liam Raynard,$^{3}$\newauthor
Rosanna~H. Tilbrook,$^{3}$
Christopher A. Watson,$^{9}$
Richard G. West$^{1,2}$ \\
$^{1}$Dept. of Physics, University of Warwick, Gibbet Hill Road, Coventry CV4 7AL, UK\\
$^{2}$Centre for Exoplanets and Habitability, University of Warwick, Gibbet Hill Road, Coventry CV4 7AL, UK\\
$^3$School of Physics and Astronomy, University of Leicester, Leicester LE1 7RH, UK\\
$^{4}$Institute of Planetary Research, German Aerospace Center, Rutherfordstrasse 2, 12489 Berlin, Germany\\
$^{5}$Astrophysics Group, Cavendish Laboratory, J.J. Thomson Avenue, Cambridge CB3 0HE, UK\\
$^{6}$Institute of Astronomy, University of Cambridge, Madingley Rise, Cambridge CB3 0HA, UK\\
$^7$Departamento de Astronom\'ia, Universidad de Chile, Camino El Observatorio 1515, Las Condes, Santiago, Chile, Casilla 36-D\\
$^8$Centro de Astrof\'isica y Tecnolog\'ias Afines (CATA), Casilla 36-D, Santiago, Chile\\
$^9$Astrophysics Research Centre, School of Mathematics and Physics, Queen's University Belfast, BT7 1NN, Belfast, UK\\
$^{10}$Instituto de Astronom\'ia, Universidad Cat\'olica del Norte, Angamos 0610, Antofagasta, 1270709, Chile\\
$^{\ddagger}$Winton Fellow
}

\date{Accepted XXX. Received YYY; in original form ZZZ}

\pubyear{2019}

\begin{document}
\label{firstpage}
\pagerange{\pageref{firstpage}--\pageref{lastpage}}
\maketitle

\begin{abstract}
We present the detection of \rewrite{high energy} white-light flares from pre-main sequence stars associated with the Orion complex, observed as part of the Next Generation Transit Survey (NGTS). With energies up to \maxflareenergy\ erg these flares are some of the most energetic white-light flare events seen to date. We have used the NGTS observations of flaring and non-flaring stars to measure the average flare occurrence rate for 4 Myr M0-M3 stars. We have also combined our results with those from previous studies to \rewrite{predict average rates for flares above \energylimit\ ergs for early M stars in nearby young associations.}

\end{abstract}


\section{Introduction}
The pre-main sequence represents the most active period in a star's lifetime, something reflected in their lightcurves. Material from surrounding discs passing in front the star can cause deep drops in their apparent flux \citep[e.g.][]{Cody14,Ansdell16}, while accretion events can cause rapid rises in flux \citep[e.g.][]{Guo18}. Another source of variability is stellar flares, explosive phenomena caused by reconnection events in the stellar magnetic field \citep[e.g.][]{Benz10}. The energy released by these reconnection events accelerates charged particles which heat the stellar photosphere, resulting in intense white-light emission.

The amplitudes and measured energies from stellar flares from pre-main sequence stars regularly dwarf those from their main sequence counterparts \citep[e.g.][]{Caramazza07,Flaccomio18}. The associated loop lengths from these flares can even dwarf the star itself, with X-ray studies measuring loop lengths of several stellar radii \citep[e.g.][]{Favata05}. If these loops link to a surrounding disc, these flares could drive accretion of disc material onto the star \citep[][]{Orlando11,Colombo19}. 

In recent years, studies of white-light flares have begun to move towards how flare properties evolve as a function of age \citep[e.g][]{Chang15,Davenport19,Ilin19}. As stars age \rewritethree{after they have reached the zero age main sequence, } they are expected to spin down and lose angular momentum through magnetic braking. Studies of open clusters and associations of known ages have shown this occurs in a mass-dependent fashion, with \rewritethree{higher mass} stars spinning down first \citep[e.g.][]{Epstein14,Matt15,Douglas19}. This spin down is expected to affect the internal dynamo also, weakening it and the magnetic field it generates. This decrease in the magnetic \rewritetwo{activity} with time has already been observed in XUV observations which probe the coronal activity \citep[e.g.][]{Jackson12}. A weakening of the magnetic activity is also expected to be reflected in a decrease in the measured energies and average occurrence rates of stellar flares. Indeed, recent studies have begun to see this in white-light flare observations. \citet{Davenport19} has studied how flare occurrence rates vary with age by combining gyrochronology with the \citet{Davenport16} \kepler\ flare sample. They were able to derive a mass and age dependent model for the flare occurrence rate, however note that their results may be affected by tidal evolution of binaries which can bias age estimates to younger ages. By using \ktwo\ observations of flares from stars within the 125 and 630 Myr Pleiades and Praesepe clusters \citet{Ilin19} measured the decrease in average flare rates between these two ages for K and M stars. They found the decrease occurs quicker than predicted from the \citet{Skumanich72} $t^{-1/2}$ law, suggesting they are still in a regime of time-variable spin down  \rewritetwo{which cannot be parameterised by a single law} \citep[e.g.][]{Barnes10}.

Measurements of average white-light flare occurrence rates so far have used clusters older than 100 Myr. Observations of systems younger than this have typically been of individual stars \citep[e.g.][]{JackmanQPP} or been X-ray observations which probe different parts of the flare reconnection model \citep[e.g.][]{Stelzer07,Flaccomio18}. During the first 100 Myr stars with masses above 0.4\Msun\ form their radiative zone \citep[][]{Baraffe15}. \citet{Gregory16} found evidence that the formation of the radiative zone is accompanied by a drop in the coronal X-ray luminosity, a parameter used as a probe of stellar magnetic activity. Studies of clusters up to 100 Myr would allow us to see whether the formation of the radiative zone has an impact on the average flaring activity. Along with this we can constrain the starting point of flare occurrence rates, informing studies of flares on the habitability of the youngest exoplanetary systems.

In this work we report the detection of white-light flares from pre-main sequence stars in the Orion Complex with the Next Generation Transit Survey (NGTS). These stars have a \rewritethree{weighted} average age of \orionage\ Myr \citep[][]{Kounkel18}. We describe how the maximum energies of these flares compare with those from main sequence stars and present measurements of the average flare occurrence rate for early M stars. We also compare our occurrence rates to those \rewritetwo{of Pleiades and Praesepe} from \citet{Ilin19} and predict upper limits on flare occurrence rates for nearby young associations.

\begin{figure}
	\includegraphics[width=\columnwidth]{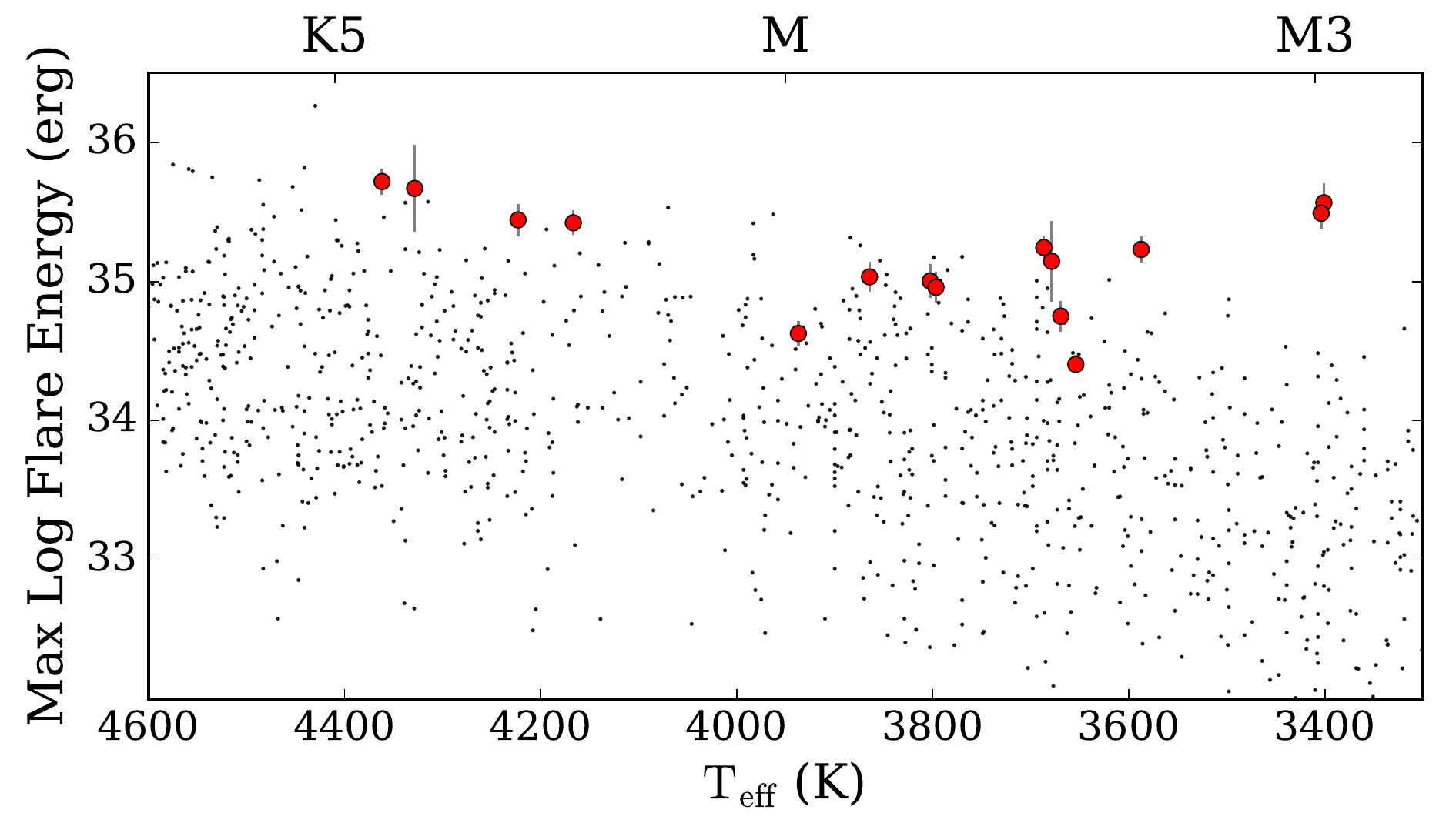}
    \caption{Maximum observed bolometric flare energy against effective temperature for our detected flaring stars. The red circles are the stars observed with NGTS. The underlying black points represent stars from \citet{Yang19}, using a posteriori corrected long cadence \kepler\ observations. Note how the young Orion-associated sources \rewritethree{detected in our survey} appear to flare with the highest observed energies. \rewritethree{We note the possibility of some stars with lower maximum flare energies in our sample which may have been missed by the detection method discussed in Sect.\,\ref{sec:obs}.}}
    \label{fig:orion_max_energy}
\end{figure}

\begin{figure} 
	\includegraphics[width=\columnwidth]{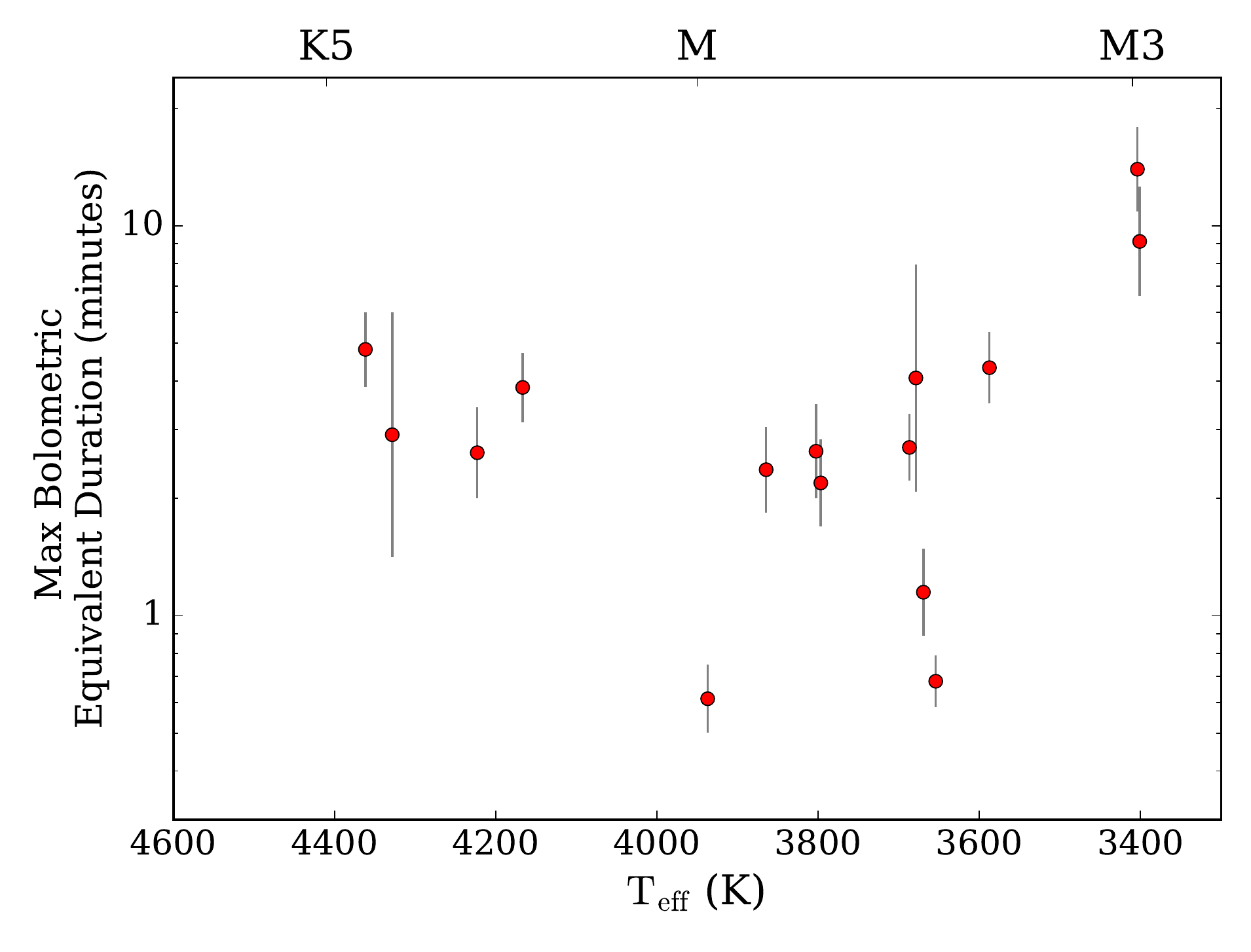}
    \caption{Maximum observed equivalent duration (in minutes) against effective temperature for our detected flaring stars. We have calculated this using our measured bolometric energies and dereddened SEDs from Sect.\,\ref{sec:sed_fit}.}
    \label{fig:orion_ed}
\end{figure}

\begin{figure}
	\centerline{\includegraphics[width=\columnwidth]{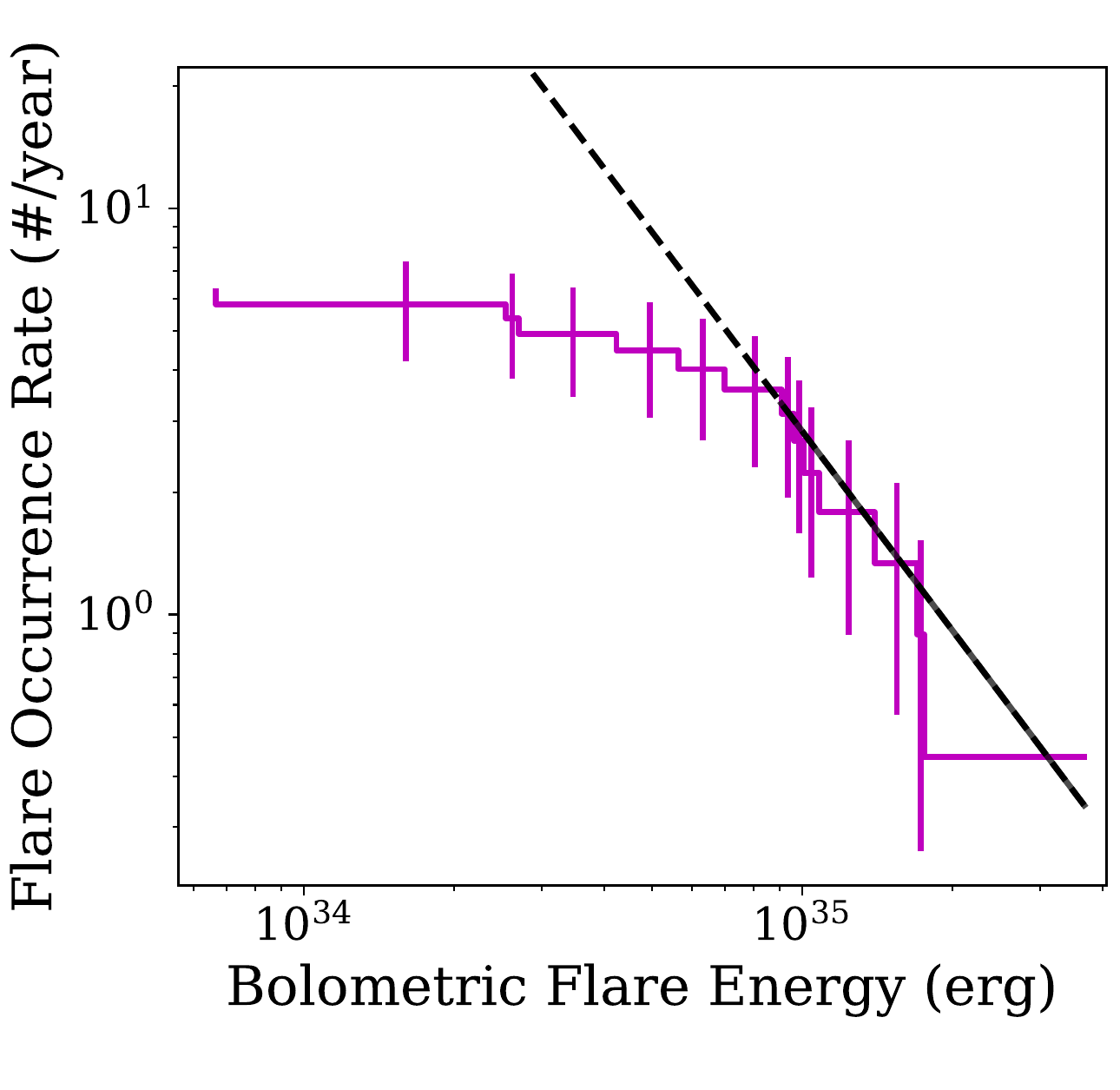}}
    \caption{Flare occurrence rate with energy for \rewrite{\orionage\ Myr M stars} associated with Orion. \rewrite{These stars were selected to have effective temperatures between 3400 and 3940K, approximately equivalent to M0-M3.} The black solid line indicates the best fitting power law in our fitting range, while the dashed line is this fit extrapolated over our measured energy range. \rewritethree{The best fitting parameters are $\alpha = 2.63\pm0.62$ ($\beta = -1.63\pm0.62$) and $C=57.4\pm21.9$.}}
    \label{fig:orion_occ_rate}
\end{figure}

\begin{table}
	\centering
	\begin{tabular}{|l|c|c|c|c|}
    \hline
    Group & $\mathrm{N_{stars,total}}$ & $\mathrm{N_{stars,flare}}$ &
    \multicolumn{1}{|p{1cm}|}{\centering $\mathrm{Age_{CMD}}$\\(Myr)} &
    \multicolumn{1}{|p{1cm}|}{\centering $\mathrm{Age_{HR}}$\\(Myr)}\tabularnewline\hline
    L1641N-1 & 31 & 7 & $3.47\pm1.89$ & $2.18\pm1.60$\tabularnewline
    onc-1 & 4 & 2 & $3.98\pm1.79$ & $4.06\pm1.48$\tabularnewline
    onc-3 & 1 &  1 & $3.31\pm1.72$ & $2.36\pm1.79$\tabularnewline
    onc-8 & 1 & 1 & $3.37\pm1.98$ & $2.99\pm2.11$\tabularnewline
    onc-11 & 1 & 0 & 3.58$\pm$1.34 & 4.85$\pm$1.97 \tabularnewline
    onc-15 & 1 & 0 & 2.51$\pm$1.86 & 2.34$\pm$1.79 \tabularnewline
    oriDS-1 & 19 & 5 & $3.02\pm1.78$ & N/A\tabularnewline
    oriDS-2 & 1 & 0 & $6.61\pm1.76$  & N/A \tabularnewline
    oriDS-3 & 7 & 1 & $5.13\pm2.14$ & N/A\tabularnewline
    oriDS-4 & 7 & 0 & $8.71\pm2.48$ & N/A \tabularnewline
    oriDS-5 & 2 & 1 & $5.01\pm1.51$ & N/A\tabularnewline
    oriDS-6 & 1 & 0 & $4.79\pm1.41$ & N/A \tabularnewline
    oriDS-7 & 5 & 1 & $6.01\pm1.65$ & N/A\tabularnewline
    oriDS-8 & 2 & 0 & $2.19\pm2.51$ & N/A\tabularnewline
	\hline
	\end{tabular}
    \caption{Identified groups for NGTS flare stars in Orion with ages from \citet{Kounkel18}. $\mathrm{N_{stars,total}}$ indicates the number of stars in our field associated with the respective group, while $\mathrm{N_{stars,flare}}$ is the number of those stars which flared. We use the $\mathrm{Age_{CMD}}$ in our analysis and list $\mathrm{Age_{HR}}$ for reference}\label{tab:orion} 
\end{table}

\renewcommand{\arraystretch}{1.25}

\begin{table*}
	\centering
	\begin{tabular}{|l|c|c|c|c|c|c|}
    \hline
    2MASS Source ID & Gaia DR2 Source ID & SpT & SpT Source & Energy (erg) & Amplitude & \thalf\ Duration (minutes)
    \tabularnewline
    \hline
2MASS J05354229-0701538  & 3016493834224711552 & G3* & SED & 
& 0.16 & 49.50 \tabularnewline
2MASS J05331830-0701552  & 3016526991372177664 & K6.5 & SED & $2.77^{+0.85}_{-0.65}\times10^{35}$ & 0.52 & 26.13 \tabularnewline
2MASS J05331830-0701552  & 3016526991372177664 & K6.5 & SED & $1.87^{+0.58}_{-0.44}\times10^{35}$ & 0.35 & 8.28 \tabularnewline
2MASS J05265619-0706548  & 3016662403101758720 & M0.5 & SED & $9.64^{+3.44}_{-2.53}\times10^{34}$ & 0.59 & 15.35 \tabularnewline
2MASS J05265619-0706548  & 3016662403101758720 & M0.5 & SED & $1.01^{+0.32}_{-0.24}\times10^{35}$ & 0.84 & 10.87 \tabularnewline
2MASS J05312392-0709304  & 3016615055382512768 & M0 & SED & $4.24^{+0.94}_{-0.77}\times10^{34}$ & 0.61 & 4.62 \tabularnewline
2MASS J05313541-0710550  & 3016614402547478784 & M1.5 & SED & $1.76^{+0.38}_{-0.31}\times10^{35}$ & 0.46 & 21.63 \tabularnewline
2MASS J05314319-0717409  & 3016424736790884736 & M1.5 & SED & $1.40^{+1.33}_{-0.68}\times10^{35}$ & 1.23 & 61.00 \tabularnewline
2MASS J05340216-0717390  & 3016505963212457088 & M3 & SED & $2.70^{+1.02}_{-0.74}\times10^{34}$ & 0.71 & 1.70 \tabularnewline
2MASS J05340216-0717390  & 3016505963212457088 & M3 & SED & $6.98^{+2.51}_{-1.85}\times10^{34}$ & 0.80 & 23.65 \tabularnewline
2MASS J05340216-0717390  & 3016505963212457088 & M3 & SED & $3.68^{+1.40}_{-1.01}\times10^{35}$ & 1.10 & 48.87 \tabularnewline
2MASS J05362219-0716064  & 3016469060853517824 & M1.5 & K18 &  & $0.44^{\mathrm{ll}}$ & 23.95 \tabularnewline
2MASS J05350988-0718045  & 3016457000585375872 & M2 & SED & $1.70^{+0.40}_{-0.32}\times10^{35}$ & 1.03 & 8.55 \tabularnewline
2MASS J05351113-0719063  & 3016456210311395456 & M1 & SED & $9.08^{+2.66}_{-2.06}\times10^{34}$ & 1.16 & 4.42 \tabularnewline
2MASS J05304799-0731323  & 3016413501156518400 & K6 & SED & $4.66^{+4.93}_{-2.40}\times10^{35}$ & 0.30 & 21.80 \tabularnewline
2MASS J05304799-0731323  & 3016413501156518400 & K6 & SED & $2.12^{+1.86}_{-0.99}\times10^{35}$ & 0.14 & 25.32 \tabularnewline
2MASS J05295834-0738069  & 3016388315468339200 & K7 & SED & $3.29^{+0.76}_{-0.62}\times10^{34}$ & 0.27 & 2.52 \tabularnewline
2MASS J05295834-0738069  & 3016388315468339200 & K7 & SED & $2.64^{+0.60}_{-0.49}\times10^{35}$ & 0.89 & 60.55 \tabularnewline
2MASS J05273635-0744059  & 3015081163645266304 & M1.5 & SED & $5.64^{+1.64}_{-1.27}\times10^{34}$ & 0.38 & 8.60 \tabularnewline
2MASS J05295183-0750306  & 3016374743371710592 & K5.5 & SED & $5.22^{+1.28}_{-1.03}\times10^{35}$ & 0.41 & 24.97 \tabularnewline
2MASS J05295183-0750306  & 3016374743371710592 & K5.5 & SED & $3.51^{+0.83}_{-0.67}\times10^{35}$ & 0.25 & 19.45 \tabularnewline
2MASS J05295183-0750306  & 3016374743371710592 & K5.5 & SED & $9.93^{+2.64}_{-2.09}\times10^{34}$ & 0.30 & 12.35 \tabularnewline
2MASS J05350942-0749194  & 3015674698061271936 & M3 & SED & $3.09^{+0.88}_{-0.68}\times10^{35}$ & 3.47 & 7.68 \tabularnewline
2MASS J05293055-0754194  & 3014873669481488896 & M1.5 & SED & $2.54^{+0.42}_{-0.36}\times10^{34}$ & 0.40 & 8.22 \tabularnewline
2MASS J05340890-0912415  & 3013742546894832384 & M0.5 & SED & $6.66^{+1.91}_{-1.49}\times10^{33}$ & 0.56 & 1.28 \tabularnewline
2MASS J05340890-0912415  & 3013742546894832384 & M0.5 & SED & $1.08^{+0.31}_{-0.24}\times10^{35}$ & 0.46 & 22.40 \tabularnewline
    
\hline
	\end{tabular}
    \caption{\rewrite{Detected flares and their measured properties. The SpT Source column indicates the method used to obtain the spectral type for this source. 2MASS J05362219-0716064 was marked in our analysis as a blended source, so does not have a measured flare energy and its amplitude is marked as a lower limit ($^\mathrm{ll}$). The spectral type of 2MASS J05354229-0701538 is marked with an asterisk for reasons discussed in Sect.\,\ref{sec:energy_discussion}. K18 indicates the spectral type is taken from \citet{Kounkel18}.}}\label{tab:flares} 
\end{table*}

\renewcommand{\arraystretch}{1.0}

\section{Observations} \label{sec:obs}
NGTS is a ground-based wide-field exoplanet survey located at the Paranal observatory in Chile. It consists of twelve 20 cm aperture telescopes with a total instantaneous field of view of 96 square degrees \citep[][]{Wheatley18}. These telescopes obtain full frame images with a 13 second cadence. The data presented in this work were collected with NGTS between the dates of 2015 Sept 23 and 2016 Apr 20, with a total of 138 nights of observation. 

In the standard mode of operations NGTS obtains lightcurves for all stars within its field of view brighter than I=16 \citep[][]{Wheatley18}. We are currently searching lightcurves from the first year of NGTS observations (Jackman et al, in prep.) for stellar flares. To search for flares we follow the \rewritetwo{two step} method described in \citet{Jackman18}. \rewritetwo{For a given star we first search individual nights for regions where there are at least 3 consecutive points above 6 median absolute deviations (MAD) from the median of the night. If there are 3 consecutive points above 6 MAD from the median, that night is automatically flagged. Secondly, we compare the median of the night against the median of the entire lightcurve for that object. If the median of the night is at least five MAD above the median lightcurve flux, then then night is automatically flagged. Nights which are raised above the median of the lightcurve are indicative of long duration and high amplitude events such as the flare discussed in \citet{JackmanQPP}. Once this automated flagging procedure was complete, we inspected each flagged night visually and removed any false positives. Such false positives can be due to astrophysical events (e.g. short period variable stars), man-made (e.g. satellites passing through our aperture) or systematics (e.g. overlapping apertures). This method is most sensitive to the highest energy flare events.}

\rewritetwo{Once we had remove any false positives from our sample, we checked the centroids of all flare candidates. Each flare is assigned to a host star, initially the star on which the aperture is placed and centered. However, if multiple stars reside within an aperture then it is possible a flare could come from a star not in the centre of the aperture. In this case, the aperture centroid will move during the flare towards the true flaring source. By checking the centroids during each flare we can confirm each source is the correct host and reassign a host star if not.}

During this flare survey we noted that one field overlapped with the southern edge of the Orion complex. This serendipitous overlap provided an opportunity to measure the white-light flaring behaviour of some of the youngest stars on the sky.

\subsection{Association}
To identify which stars are associated with groups within Orion we crossmatched all NGTS sources in the Orion-overlapping field with the catalogue of \citet{Kounkel18}. Using \Gaia\ DR2 and APOGEE data, \citet{Kounkel18} have mapped distinct groups within the Orion complex and their catalogue provides designations for Orion-associated stars and field stars. From this matching we identified 83 Orion-associated stars which fell within our field of view. Of these, 19 flared at least once during our observations. \rewrite{The smallest and largest amplitude detected flares are shown in Fig.\,\ref{fig:smallest_flare} and Fig.\,\ref{fig:largest_flare}.} The groups and ages of the stars in our sample are shown in Tab.\,\ref{tab:orion}. We can see in Tab.\,\ref{tab:orion} that these groups have similar ages. Consequently, we have considered them as a single group of pre-main sequence stars with an average age of \orionage\ Myr for the rest of this analysis. We calculated this age \rewritethree{and uncertainty by taking a weighted average of the $\mathrm{Age_{CMD}}$ values presented in Tab.\,\ref{tab:orion}.} \rewritethree{The  $\mathrm{Age_{CMD}}$ were calculated by \citet{Kounkel18} using colour-magnitude diagrams used APOGEE and \gaia\ DR2 colours with PARSEC models to obtain absolute magnitudes for all young stars in their sample, including for young stellar objects without reliable parallaxes. The $\mathrm{Age_{HR}}$ values listed in Tab.\,\ref{tab:orion} used the \gaia\ parallaxes and APOGEE measured effective temperatures to obtain luminosities, which were compared with PARSEC models to infer ages. The dependence on the effective temperature results in a smaller sample having $\mathrm{Age_{HR}}$ values, as seen in Tab.\,\ref{tab:orion}. Due to the lack of $\mathrm{Age_{HR}}$ values for all stars in our sample, we have used the $\mathrm{Age_{CMD}}$ in this work.}

\subsection{Stellar Properties} \label{sec:sed_fit}
Eight sources in our sample had effective temperatures and angular sizes measured by \citet{Kounkel18} from APOGEE spectra. We obtained the stellar radius for these sources by combining the angular sizes with their measured distances from the \citet{BailerJones18} catalogue, using parallaxes from \Gaia\ DR2 \citep{Gaia2016,GaiaDR2}. 

For the rest of our sources we performed Spectral Energy Distribution (SED) fitting using the PHOENIX v2 spectral library \citep[][]{Allard12}. Each source was crossmatched with \gaia\ DR2, 2MASS \citep{2MASS_2006}, APASS \citep{APASS_14}, SDSS \citep{Aguado18} and ALLWISE \citep{ALLWISE2014} to obtain broadband photometry. Distance information was obtained from the catalogue of \citet{BailerJones18}. To account for the effects of reddening and extinction we fit for $A_{V}$, with $R_{V}=3.1$ \citep[][]{Cardelli89}. \rewritetwo{$R_{V}$ is the ratio of the total to selective extinction.} We applied a Gaussian prior on the reddening based on the values measured from APOGEE spectra for the groups in Tab.\,\ref{tab:orion} by \citet{Kounkel18}. This was chosen over existing dust maps of \citet{Green19} as we found \rewritetwo{half of the stars in our sample had poor likelihoods in their dust map fits, using the criteria from \citet{Green19}.}

\rewrite{To avoid the effects of blending between close neighbours influencing our SED fitting, we checked all sources for stars within 15\arcsec\ (the radius of an NGTS aperture) which might contribute significant levels of flux to the NGTS lightcurve. Sources which had nearby neighbours were flagged and did not have SED fits generated. Consequently, these stars (flaring and non-flaring) were not used in our analysis.}

From our \rewrite{analysis} we obtain \kdwarfnumber\ K and \mdwarfnumber\ M stars in our sample. The rest of the sample either had effective temperatures hotter than K spectral type, or did not have a reliable SED fit. 

\subsection{Flare Properties} \label{sec:energy_method}
To calculate flare energies we follow an \rewritetwo{adjusted} version of the method from \cite{Shibayama13}. \rewrite{The original method uses the observed flare amplitude to normalise an assumed 9000$\pm$500\,K flare blackbody \citep[e.g.][]{Hawley1992} relative to the quiescent stellar spectrum within the observed bandpass. The bolometric flare energy is measured by integrating over the renormalised 9000\,K flare spectrum at each observed time. We have modified this method in our analysis to account for the effects of reddening on the observed flare and stellar spectrum \citep[e.g.][]{PaudelBD}, using the fitted values of $A_{V}$ from Sect.\,\ref{sec:sed_fit}.} \rewritetwo{We used the average Milky Way exinction model from \citet{Cardelli89} in this method.}

When applying this method we assumed the underlying \rewrite{quiescent} stellar flux could be modelled with a linear baseline between the start and end of the flare. For the stellar properties (to generate the quiescent stellar spectrum) we use the radii and  effective temperatures from Sect.\,\ref{sec:sed_fit}. 

\rewritetwo{We calculated the \thalf\ duration and amplitude for all flares. These values are listed in Tab.\,\ref{tab:flares}. The \thalf\ duration is a commonly used measure of flare duration \citep[e.g.][]{Davenport2014} and is defined as the time spent above half of the flare amplitude. We calculated the amplitude following the method from \citet{Hawley14} using $\frac{\Delta F}{F} = \frac{F - F_{o}}{F_{o}}$ where $F_{o}$ is the out-of-flare flux. The value of $F_{o}$ is calculated from the median of the flux preceding the start of the flare. As these parameters are independent of spectral type we have listed both \thalf\ and the amplitude for all flares in Tab.\,\ref{tab:flares}.}

\section{Results and Discussion} \label{sec:energy_discussion}
From an analysis of 83 stars associated with the Orion complex which were observed with NGTS we detected \flarenumber\ flares from 17 stars. 16 of these 17 flare stars had effective temperatures from our SED fits or the \citet{Kounkel18} sample, which corresponded to spectral types between \rewrite{G3} and M3. Twelve stars were M spectral type and four were K. 
The observed flares had energies ranging between \minflareenergy\ and \maxflareenergy\ ergs. 

\rewrite{The presence of a flare from a G3 star in our sample is interesting due to the rarer nature of flares from these stars. However, investigating this source further shows it has previously been classified as K4 from low-resolution spectra by \citet{Pravdo95} and as G8 from spectroscopic observations by \citet{Strom90}. Based on these observations it is likely this source is later in spectral type than our SED fitting suggests, however without an exact classification we have chosen to leave it out from our energy analysis.} \rewritetwo{However, we still list the NGTS flare amplitude and duration for this star in Tab.\,\ref{tab:flares}, as these do not depend on the stellar properties.}

\subsection{Maximum Flare Energy}

In Fig.\,\ref{fig:orion_max_energy} we have compared the maximum flare energy of the Orion-associated stars to main sequence stars observed with \kepler\ from \citet{Yang19}. The Orion-associated stars reside at the top of the energy envelope with maximum flare energies between \minmaxflareenergy\ and \maxflareenergy\ erg. These energies are similar to those observed by \citet{Flaccomio18} from simultaneous mIR, optical and X-ray observations of pre-main sequence stars in the $\sim$3 Myr old NGC 2264 star forming region. Adopting the ratio between optical and X-ray energies (0.5-8.0 keV) from \citet{Flaccomio18} \rewritetwo{we estimate that }our maximum X-ray flare energies are between \minmaxxrayflareenergy\ and \maxxrayflareenergy\ erg.

\rewritetwo{Our measured bolometric flare energies, while similar to the \citet{Flaccomio18} sample, reside towards the lower end of their measured energies. This is possibly due the \citet{Flaccomio18} sample observing more stars in their study, approximately 500 stars for 60 days with \textit{CoRoT}.  Consequently, this suggests that while our flares have high energies, they are not at the maximum limit of flare energies for these young stars. Further observations are therefore required to reach the maximum flare energy limit for 4 Myr old stars.}

We have also measured the bolometric equivalent durations (EDs) for each flare in our sample, using the bolometric flare energies and our SED fits from Sect.\,\ref{sec:sed_fit}. The ED is the amount of time required for the quiescent luminosity of the star to emit an energy equal to an observed flare \citep[e.g.][]{Hawley14}. It is usually expressed in units of time. We calculated our EDs by dividing the bolometric flare energies by the stellar bolometric luminosity. We calculated the bolometric luminosity by integrating in wavelength over our best fitting SED fits (after removing the effect of extinction) and multiplying by $4\pi d^2$ where $d$ is the distance. Fig.\,\ref{fig:orion_ed} shows the maximum bolometric ED for each star in our sample, showing that the equivalent durations have values up to 15 minutes, however the majority seem to have values of a few minutes. These values are lower than those reported for rare extreme flare events from other pre-main sequence stars \citep[e.g.][]{Paudel18,JackmanQPP} which can stretch to hours \citep[although the ED is often reported for a single filter, which can change measured values, e.g.][]{Hawley14}. This again suggests that our sample, while exhibiting high flare energies, has not reached the maximum flare energy limit for these young stars, something which will be probed in future studies. 

\subsection{Flare Occurrence Rate of M stars} \label{sec:m_star_rates}
\rewritetwo{Previous studies \citep[e.g.][]{Lacy76,Hilton11,Hawley14} have shown that flares occur with a power law distribution in energy. This is typically written as 
\begin{equation}
    dN(E) \propto  E^{-\alpha}dE
\end{equation}
where $E$ is the flare energy and $\alpha$ is the power law index. From this, a linear relation for the cumulative flare frequency distribution (CFFD) can be written as
\begin{equation}
    \log{\nu} = C + \beta \log{E}
\end{equation}
where $\nu$ is the cumulative flare frequency, $C$ is a normalisation constant and $\beta = 1 - \alpha$ \citep[][]{Hawley14}. The value of $\alpha$ dictates how often high energy flares occur relative to those of lower energy and has been observed to vary across different spectral types. For example, the Sun has $\alpha\approx$1.75 \citep[e.g.][]{Crosby93,Shimizu95,Aschwanden2000}, while ultracool dwarfs have had measured $\alpha$ values from 1.4 to 2 \citep[e.g.][]{Paudel18}.}

In Fig.\,\ref{fig:orion_occ_rate} we have plotted the average flare occurrence rate for stars in our Orion-associated sample. This was done for stars with effective temperatures between 3400 and 3940\,K, representative of early-M (M0-M3) in spectral type \citep[e.g.][]{Pecaut13}. \rewrite{To fit the power law shown in Fig.\,\ref{fig:orion_occ_rate} we have used the Python {\scshape powerlaw} package \citep[][]{Alstott14}. {\scshape powerlaw} is designed for fitting to heavy-tailed distributions and has been used previously for analysis of flare occurrence rates \citep[e.g.][]{Lin19}. }

\rewritetwo{To make sure our sample is complete in terms of the detected flare energies we performed flare injection and retrieval tests, following a similar methodology to those employed by \citet{Davenport16} and \citet{Ilin19}. We first injected 1000 artificial flares into the lightcurves of every M star in our sample. These flares were generated using the \citet{Davenport2014} empirical flare model. The amplitude of each flare was chosen randomly to have a value from between 0.01 and 3.5 times the median of an object's lightcurve. The \thalf\ timescale for each flare was chosen randomly from between 30 seconds and 70 minutes, to encompass the full range of detected flares. The automated detection method outlined in Sect.\,\ref{sec:obs} was then run for each object on every night with an injected flare. We calculated the energies of each injected flare on a star-by-star basis, applying the energy calculation from Sect.\,\ref{sec:energy_method} to the artificially generated flare. Injected flares were given a flag of 1 if they were automatically detected by our method, 0 if not.}

\rewritetwo{The fraction of recovered flares was calculated as a function of energy for each star, using 20 bins in energy. Following \citet{Davenport16}, the recovery fraction was smoothed using a Wiener filter of three bins, to smooth any jumps and drops between bins. For each star we measured the flare energy at which at least 68 per cent of flares were retrieved, which was recorded for each star, as in \citet{Davenport16}.}

\rewritetwo{To ensure a complete sample when calculating the flare occurrence rate and when fitting the associated power law, an energy limit was used. For a given energy limit, only stars with a 68 per cent recovery fraction energy below this limit were used in calculating the average flare occurrence rate. This applied whether or not a star had any detected flares within its NGTS lightcurve. The power law itself, as seen in Fig.\,\ref{fig:orion_occ_rate}, was fit only to flares (from stars within this complete sample) with energies above the chosen energy limit. To select an energy limit, we tested a series of values to maximise the number of flares used in our power law fitting, without compromising the total number of stars used. Low energy limit values would exclude most stars (and their flares) from use in the flare occurrence rate fitting. Very high energy limit values, while including all stars in the sample, would have very few flares to fit the power law to. We found a value of $9\times10^{34}$ erg best satisfied this compromise and was thus used as our energy limit.}

\rewritetwo{29 stars in our 3400-3940\,K sample (40 stars in total) had 68 per cent recovery fraction energies below our limit of $9\times10^{34}$ erg, and were used in our occurrence rate analysis. 10 of these stars had a detected flare. 8 flares (from 7 stars) had energies above the $9\times10^{34}$ erg, to which we fitted a power law distribution. 6 flares (from 5 stars) had energies below this threshold and were not used in our power law fitting. Some stars in our sample flared more than once, with energies above and below $9\times10^{34}$ erg.} \rewritethree{The full sample of flares detected from stars in the 3400-3940\,K sample, along with the 68 per cent completeness energy for each star, are presented in Tab.\,\ref{tab:app_early_m_flares}. The full sample of completeness energies for each individual star, flaring and non-flaring, in the 3400-3940\,K sample is presented in Tab.\,\ref{tab:full_early_m_stars}.}

\rewritetwo{From fitting our sample above this threshold we measured \rewritetwo{$\alpha = 2.63\pm0.62$ \rewritetwo{($\beta = -1.63\pm0.62$)} and $C=57.4\pm21.9$}.}  
An $\alpha$ value greater than 2 indicates that lower energy flares dominate the total flare energy distribution. \rewritetwo{$\alpha > 2$ has also been noted in previous works as a requirement for low energy flares to heat Solar and stellar coronas \citep[e.g.][]{Doyle85,Parker88,Gudel03}. With our sample we are not able to confidently rule out high energy flares dominating the distribution, although we note the possibility that flares may be a significant contributor to the quiescent magnetic environment of young M stars. This value is also consistent with the value of $\alpha=2.2\pm0.2$ measured by \citet{Caramazza07} for X-ray flares from Orion associated 0.1-0.3\,\Msun\ stars.}

From our power law fit in Fig.\,\ref{fig:orion_occ_rate} we estimate that the average 3400-3940\,K star in the Orion complex flares with an energy above $10^{35}$ erg once every \rewrite{130} days. This is almost \rewrite{1000} times the rate measured by \citet{Lin19} for M dwarfs using \ktwo\ observations, highlighting how active these \rewritetwo{pre-main sequence} sources are.

Previous studies have noted there may be a break off and steepening in the power law towards the highest flare energies \citep[e.g.][]{Chang15}. We note the possibility that this is the regime we have measured and that $\alpha$ at lower energies may be less than what we have measured. \rewrite{More observations of white-light flares from Orion-associated stars will allow us to further constrain $\alpha$ at the highest energies, by increasing the number of stars and rare high energy flares observed.}

\subsubsection{Comparison with other clusters}
\rewrite{\citet{Ilin19} found from \ktwo\ observations that between the 125 Myr Pleiades (M44) and 630 Myr Praesepe (M45) open clusters the average flare occurrence rate (for bolometric energies above $3.4\times10^{33}$ erg) of early-M stars decreases quicker than the Skumanich $t^{-1/2}$ law. They found that the activity of hotter stars drops the quickest, tying in with previous observations of hotter stars spinning down more quickly than their cooler counterparts \citep[e.g.][]{Amard19}. The rate of this decrease, if it tracks with the change in surface rotation, is also expected to be variable at young ages \citep[e.g.][]{Barnes10}. We can use our sample to try and extend this work for the highest energy flares down to stars of 4 Myr in age.}

\rewrite{To measure the change in the flare rate for energies above \energylimit\ erg we have split our sample into the 3500-3750\,K and 3750-4000\,K groups from \citet{Ilin19}. In each subset we have measured the rates directly by linearly interpolating the measured average occurrence rates to the value at \energylimit\ erg. To calculate the rates for the Pleiades and Praesepe we first \rewritetwo{multiplied} the \kepler\ bandpass occurrence rates measured by \citet{Ilin19} by 3.1 to convert them to bolometric energies, using the conversion factor from \citet{Paudel18}. We then extrapolated these power laws to \energylimit\ erg.}

For the 3500-3750\,K sample the \rewritethree{measured average occurrence rate of flares above $1\times10^{35}$ ergs changes from 2.0 to $7.3\times10^{-2}$ and then to $6.5\times10^{-2}$ flares per year from Orion, through the Pleiades, to Praesepe. The change in this measured rate with age consequently goes from $-1.6\times10^{-2}$ to $-1.73\times10^{-5}$ flares/year/Myr. For the 3750-4000\,K sample the rate of flares above $1\times10^{35}$ ergs changes from 2.3 to $2.5\times10^{-1}$ and then to $3.5\times10^{-2}$. The change in this measured rate with age consequently goes from $-1.7\times10^{-2}$ to $-4.22\times10^{-4}$ flares/year/Myr. We can see that, from the change in the measured occurrence rate between open clusters, the observed activity of hotter stars (above $1\times10^{35}$ ergs) is dropping at a faster rate in the first 600 Myr than cooler, lower mass, stars.}

We also note that the \rewrite{change of occurrence rate of flares above \energylimit\ ergs is not linear in log-log space, }
in disagreement with the model derived by \citet{Davenport19}. However, as noted in Sect.\,\ref{sec:m_star_rates} 
flaring behaviour may change at the energies we are probing. 
These changes \citep[e.g. breaks in power laws;][]{Chang15} were not included in the \citet{Davenport19} model and as such a strict comparison may not be valid. \rewritetwo{We also note that the number of flares in our 3500-3750\,K and 3750-4000\,K Orion subsets are 5 and 6 respectively. The small sample size in these subsets combined with the extrapolation of the \citet{Ilin19} occurrence rates means the values in Tab.\,\ref{tab:predict} should be used with caution and best as estimates.}

By combining these datasets together and linearly interpolating in age we can \rewritetwo{estimate the} average flare rates for M stars in other clusters and associations. Some predictions for nearby young clusters are given in Tab.\,\ref{tab:predict} for E>\energylimit\ ergs. We find that up to 70 Myr both samples can flare up to once a year with an energy above \energylimit\ ergs.

\begin{table*}
	\centering
	\begin{tabular}{|l|c|c|c|c|c|}
    \hline
    Cluster &  \multicolumn{1}{|p{1cm}|}{\centering Age\\ (Myr)} &  \multicolumn{1}{|p{2cm}|}{\centering 3500-3750\,K\\ (\#/year)} & $\mathrm{n_{flares}}$ & 
     \multicolumn{1}{|p{2cm}|}{\centering 3750-4000\,K\\ (\#/year)} & $\mathrm{n_{flares}}$ \tabularnewline \hline
    \rewrite{Orion*} & 4 & 2.0 & 5 & 2.3 & 6 \tabularnewline
     \rewrite{Beta Pic} & 24 & 1.7 & \rewritethree{c} & 2.0 & \rewritethree{c}  \tabularnewline
     \rewrite{Carina} & 45 & 1.3 & \rewritethree{c} & 1.6 & \rewritethree{c} \tabularnewline
     \rewrite{Pleiades} $\ddagger$ & 125 & $7.3\times10^{-2}$ & e &  2.5$\times10^{-1}$ & e \tabularnewline
     \rewrite{AB Doradus} & 149 & 7.3$\times10^{-2}$ & e & 2.4$\times10^{-1}$ & e \tabularnewline
     \rewrite{Praesepe} $\ddagger$ & 630 & 6.5$\times10^{-2}$ & e & 3.5$\times10^{-2}$ & e \tabularnewline
	\hline
	\end{tabular}
    \caption{\rewrite{Predictions of }the average flare occurrence rate for some nearby young associations, for flare energies above \energylimit\ erg. We have calculated these by combining our measured occurrence rates with those \rewritetwo{extrapolated }from \citet{Ilin19}. For intermediate age clusters we \rewritetwo{linearly interpolated between measured samples.} The * indicates these are our measured values for Orion. \rewrite{The $\ddagger$ indicates these are the \citet{Ilin19} \rewritetwo{extrapolated} values for Pleiades and Praesepe, again shown for reference.} \rewritetwo{Also shown are the number of flares in each sample. \rewritethree{``c'' indicates a predicted value which has been calculated from the Orion, Pleiades and Praesepe samples.} ``e'' indicates an estimate uses the extrapolated \citet{Ilin19} occurrence rates.} Quoted ages for groups other than Orion, \rewrite{Pleiades and Praesepe} come from \citet{Bell15}.}\label{tab:predict} 
\end{table*}

\subsubsection{Formation of the radiative zone}

With an age of 4 Myr, the pre-main sequence early M stars in Orion are fully convective. The magnetic fields of pre-main sequence early-M stars have previously been characterised as being large scale and axisymmetric. As these early M stars evolve they will contract and spin up, something which seems counter-intuitive to a decrease in flare rates. However, during this process they begin to form an inner radiative zone which is separated from the outer convective regions by a tachocline. From the \citet{Baraffe15} models this radiative zone reaches its maximum size at roughly 100 Myr. The formation of the radiative zone is expected to be accompanied by a change in the stellar magnetic topology and has previously been linked to observed decreases in the coronal X-ray luminosity \citep[][]{Gregory16}, which is often used as a tracer of magnetic activity \citep[e.g.][]{Wright11}. The transition between fully and partially convective interiors for main sequence stars may also be linked to a decrease in the magnetic field strength, something previously noted by \citet{Mullan18Tr1} \citep[see Fig. 14 of][]{Moutou17}. It may be that the formation of a radiative zone for early M stars causes a drop in flare energies and rates between 4 and 125 Myr. After 125 Myr the average flare behaviour would more closely follow the decay of stellar rotation. Further observations of stars in Orion to confirm our measured average occurrence rates along with further observations of open clusters or moving groups with intermediate ages will be needed to test this hypothesis for early M stars \citep[e.g. the 30-40 Myr Octans group;][]{Murphy15} and other spectral types.

\section{Conclusions}
In this work we have presented the detection of \rewrite{high energy} white-light stellar flares from \orionage\ Myr pre-main sequence stars associated with the Orion Complex. These are some of the youngest stars to have a measured white-light flare occurrence rate. By combining our measured occurrence rates with those in the literature we identified that the \rewrite{rate of change} in the average flare occurrence rate \rewrite{above \energylimit\ ergs} for M dwarfs is variable with time, \rewritetwo{appearing to change rapidly during the 100 Myr before slowing.} This result is similar to findings for surface rotation. We have used these rates to \rewrite{predict the flare occurrence rates} for nearby moving groups.

\section*{Acknowledgements}
\rewritetwo{We thank the referee for their help in greatly improving this manuscript.} This research is based on data collected under the NGTS project at the ESO La Silla Paranal Observatory. The NGTS facility is funded by a consortium of institutes consisting of 
the University of Warwick,
the University of Leicester,
Queen's University Belfast,
the University of Geneva,
the Deutsches Zentrum f\" ur Luft- und Raumfahrt e.V. (DLR; under the `Gro\ss investition GI-NGTS'),
the University of Cambridge, together with the UK Science and Technology Facilities Council (STFC; project reference ST/M001962/1). 
JAGJ is supported by STFC PhD studentship 1763096.
PJW is supported by STFC consolidated grant ST/P000495/1.
EG gratefully acknowledges support from the David and Claudia Harding Foundation in the form of a Winton Exoplanet Fellowship.

This publication makes use of data products from the Two Micron All Sky Survey, which is a joint project of the University of Massachusetts and the Infrared Processing and Analysis Center/California Institute of Technology, funded by the National Aeronautics and Space Administration and the National Science Foundation. This publication makes use of data products from the Wide-field Infrared Survey Explorer, which is a joint project of the University of California, Los Angeles, and the Jet Propulsion Laboratory/California Institute of Technology, funded by the National Aeronautics and Space Administration. This work has made use of data from the European Space Agency (ESA) mission {\it Gaia} (\url{https://www.cosmos.esa.int/gaia}), processed by the {\it Gaia} Data Processing and Analysis Consortium (DPAC, \url{https://www.cosmos.esa.int/web/gaia/dpac/consortium}). Funding for the DPAC has been provided by national institutions, in particular the institutions participating in the {\it Gaia} Multilateral Agreement.

\section*{Data Availability}
Stellar and flare properties are available upon request to JAGJ.

\bibliographystyle{mnras}
\bibliography{references}

\appendix
\section{The smallest and largest detected flares}
\begin{figure}
	\includegraphics[width=\columnwidth]{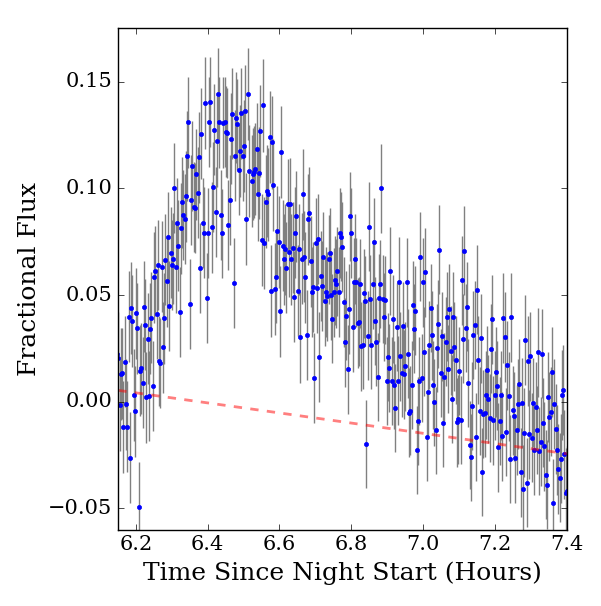}
    \caption{The lowest amplitude flare detected in our sample. The red dashed line is an example of the baseline used when calculating flare energies, as discussed in Sect.\,\ref{sec:energy_method}}.
    \label{fig:smallest_flare}
\end{figure}

\begin{figure}
	\includegraphics[width=\columnwidth]{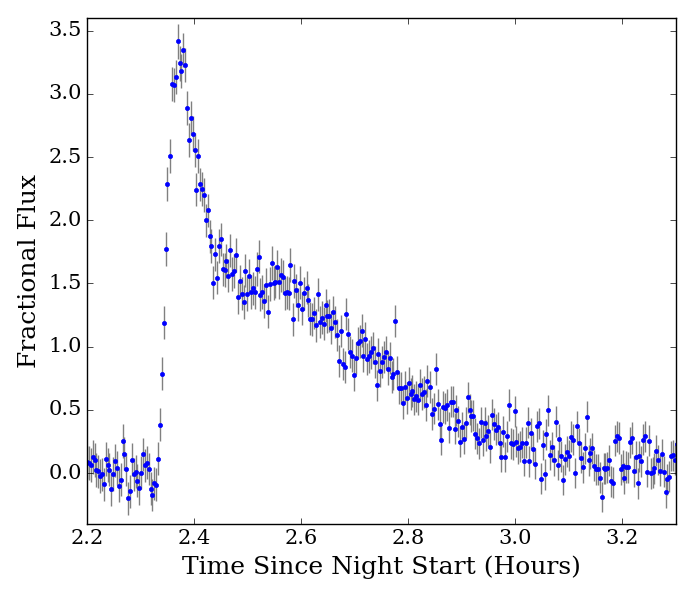}
    \caption{The highest amplitude flare detected in our sample.}
    \label{fig:largest_flare}
\end{figure}

\section{Early M stars and their completeness energies}

\renewcommand{\arraystretch}{1.25}

\begin{table*}
	\centering
	\begin{tabular}{|l|c|c|c|c|c|c|}
    \hline
    2MASS Source ID & Gaia DR2 Source ID & SpT & SpT Source & Flare Energy (erg) & Completeness Energy (erg) & Used?
    \tabularnewline
    \hline
2MASS J05265619-0706548  & 3016662403101758720 & M0.5 & SED & $9.64^{+3.44}_{-2.53}\times10^{34}$ & $7.26\times 10^{34}$ & Y \tabularnewline
2MASS J05265619-0706548  & 3016662403101758720 & M0.5 & SED & $1.01^{+0.32}_{-0.24}\times10^{35}$ & $7.26\times 10^{34}$ & Y \tabularnewline
2MASS J05312392-0709304  & 3016615055382512768 & M0 & SED & $4.24^{+0.94}_{-0.77}\times10^{34}$ & $6.42\times 10^{34}$ & Y \tabularnewline
2MASS J05313541-0710550  & 3016614402547478784 & M1.5 & SED & $1.76^{+0.38}_{-0.31}\times10^{35}$ & $5.52\times 10^{34}$ & Y \tabularnewline
2MASS J05314319-0717409  & 3016424736790884736 & M1.5 & SED & $1.40^{+1.33}_{-0.68}\times10^{35}$ & $4.38\times 10^{34}$ & Y \tabularnewline
2MASS J05340216-0717390  & 3016505963212457088 & M3 & SED & $2.70^{+1.02}_{-0.74}\times10^{34}$ & $6.51\times 10^{34}$ & Y \tabularnewline
2MASS J05340216-0717390  & 3016505963212457088 & M3 & SED & $6.98^{+2.51}_{-1.85}\times10^{34}$ & $6.51\times 10^{34}$ & Y \tabularnewline
2MASS J05340216-0717390  & 3016505963212457088 & M3 & SED & $3.68^{+1.40}_{-1.01}\times10^{35}$ & $6.51\times 10^{34}$ & Y \tabularnewline
2MASS J05350988-0718045  & 3016457000585375872 & M2 & SED & $1.70^{+0.40}_{-0.32}\times10^{35}$ & $6.79\times 10^{34}$ & Y \tabularnewline
2MASS J05351113-0719063  & 3016456210311395456 & M1 & SED & $9.08^{+2.66}_{-2.06}\times10^{34}$ & $7.83\times 10^{34}$ & Y \tabularnewline
2MASS J05273635-0744059  & 3015081163645266304 & M1.5 & SED & $5.64^{+1.64}_{-1.27}\times10^{34}$ & $6.09\times 10^{34}$ & Y \tabularnewline
2MASS J05350942-0749194  & 3015674698061271936 & M3 & SED & $3.09^{+0.88}_{-0.68}\times10^{35}$ & $1.58\times 10^{35}$ & N \tabularnewline
2MASS J05293055-0754194  & 3014873669481488896 & M1.5 & SED & $2.54^{+0.42}_{-0.36}\times10^{34}$ & $6.64\times 10^{34}$ & Y \tabularnewline
2MASS J05340890-0912415  & 3013742546894832384 & M0.5 & SED & $6.66^{+1.91}_{-1.49}\times10^{33}$ & $6.03\times 10^{34}$ & Y \tabularnewline
2MASS J05340890-0912415  & 3013742546894832384 & M0.5 & SED & $1.08^{+0.31}_{-0.24}\times10^{35}$ & $6.03\times 10^{34}$ & Y \tabularnewline

\hline
	\end{tabular}
    \caption{\rewritethree{Detected flares, their energies and the 68\% completeness energies for the stars in the full 3400-3940\,K sample. The final column indicates whether this star was used in our calculatation of the flare occurrence rate in Sect.\,\ref{sec:m_star_rates}.} }\label{tab:app_early_m_flares} 
\end{table*}

\begin{table*}
	\centering
	\begin{tabular}{|l|c|c|c|c|c|c|}
    \hline
    2MASS Source ID & Gaia DR2 Source ID & SpT & SpT Source & Completeness Energy (erg) & Used?
    \tabularnewline
    \hline
    2MASS J05265619-0706548  & 3016662403101758720 & M0.5 & SED & $7.26\times 10^{34}$ & Y \tabularnewline
    2MASS J05344294-0703497  & 3016533996462888192 & M0.5 & SED & $7.26\times 10^{34}$ & Y \tabularnewline
    2MASS J05302813-0706421  & 3016629417753191424 & M1.5 & SED & $3.64\times 10^{34}$ & Y \tabularnewline
    2MASS J05362055-0705317  & 3016488955142025984 & M0 & KOUNKEL & $1.11\times 10^{35}$ & N \tabularnewline
    2MASS J05313128-0708564  & 3016615261540932224 & M3 & SED & $1.35\times 10^{35}$ & N \tabularnewline
    2MASS J05312392-0709304  & 3016615055382512768 & M0 & SED & $6.42\times 10^{34}$ & Y \tabularnewline
    2MASS J05313541-0710550  & 3016614402547478784 & M1.5 & SED & $5.52\times 10^{34}$ & Y \tabularnewline
    2MASS J05335093-0711038  & 3016510911014768512 & M1.5 & SED & $6.52\times 10^{34}$ & Y \tabularnewline
    2MASS J05343263-0715180  & 3016460883235690624 & M3 & SED & $7.47\times 10^{34}$ & Y \tabularnewline
    2MASS J05314319-0717409  & 3016424736790884736 & M1.5 & SED & $4.38\times 10^{34}$ & Y \tabularnewline
    2MASS J05290015-0719465  & 3016598837585920512 & M1 & SED & $5.99\times 10^{34}$ & Y \tabularnewline
    2MASS J05340216-0717390  & 3016505963212457088 & M3 & SED & $6.51\times 10^{34}$ & Y \tabularnewline
    2MASS J05350988-0718045  & 3016457000585375872 & M2 & SED & $6.79\times 10^{34}$ & Y \tabularnewline
    2MASS J05313990-0719450  & 3016423053163711488 & M2 & SED & $7.34\times 10^{34}$ & Y \tabularnewline
    2MASS J05325564-0720064  & 3016501874403636480 & M3 & SED & $1.46\times 10^{35}$ & N \tabularnewline
    2MASS J05351113-0719063  & 3016456210311395456 & M1 & SED & $7.83\times 10^{34}$ & Y \tabularnewline
    2MASS J05351686-0719019  & 3016455969793224192 & M1 & SED & $8.38\times 10^{34}$ & Y \tabularnewline
    2MASS J05351770-0720152  & 3016455866714010880 & M1 & SED & $9.94\times 10^{34}$ & N \tabularnewline
    2MASS J05305836-0722365  & 3016417212008224896 & M2 & SED & $4.66\times 10^{34}$ & Y \tabularnewline
    2MASS J05291038-0724262  & 3016594576978615936 & M3 & SED & $6.76\times 10^{34}$ & Y \tabularnewline
    2MASS J05322860-0724037  & 3016408141037780736 & M3 & SED & $4.54\times 10^{34}$ & Y \tabularnewline
    2MASS J05341974-0720320  & 3016457893938475904 & M2 & SED & $7.38\times 10^{34}$ & Y \tabularnewline
    2MASS J05344904-0726070  & 3016453152294702848 & M2 & SED & $5.70\times 10^{34}$ & Y \tabularnewline
    2MASS J05332977-0726348  & 3016450571017334016 & M1.5 & KOUNKEL & $3.68\times 10^{35}$ & N \tabularnewline
    2MASS J05355628-0728317  & 3016438549405961472 & M1.5 & SED & $5.14\times 10^{34}$ & Y \tabularnewline
    2MASS J05295064-0734543  & 3016576366317290240 & M1 & SED & $5.39\times 10^{34}$ & Y \tabularnewline
    2MASS J05273635-0744059  & 3015081163645266304 & M1.5 & SED & $6.09\times 10^{34}$ & Y \tabularnewline
    2MASS J05320092-0745170  & 3016344644241021824 & M2.5 & SED & $1.24\times 10^{35}$ & N \tabularnewline
    2MASS J05293238-0747456  & 3016378832180580352 & M2 & SED & $1.50\times 10^{35}$ & N \tabularnewline
    2MASS J05350942-0749194  & 3015674698061271936 & M3 & SED & $1.58\times 10^{35}$ & N \tabularnewline
    2MASS J05293055-0754194  & 3014873669481488896 & M1.5 & SED & $6.64\times 10^{34}$ & Y \tabularnewline
    2MASS J05343717-0757402  & 3015575398418115456 & M0.5 & KOUNKEL & $1.08\times 10^{35}$ & N \tabularnewline
    2MASS J05361957-0759448  & 3015653532462547200 & M2 & SED & $1.89\times 10^{35}$ & N \tabularnewline
    2MASS J05304518-0804494  & 3016309116271690240 & M2 & SED & $4.89\times 10^{34}$ & Y \tabularnewline
    2MASS J05313203-0809001  & 3016305645938094592 & M2.5 & SED & $5.67\times 10^{34}$ & Y \tabularnewline
    2MASS J05344653-0818539  & 3015547807548427264 & M2.5 & SED & $1.60\times 10^{35}$ & N \tabularnewline
    2MASS J05281520-0844070  & 3014757258688194048 & M2 & SED & $7.26\times 10^{34}$ & Y \tabularnewline
    2MASS J05371561-0859214  & 3015242177675724032 & M0 & SED & $7.34\times 10^{34}$ & Y \tabularnewline
    2MASS J05340890-0912415  & 3013742546894832384 & M0.5 & SED & $6.03\times 10^{34}$ & Y \tabularnewline
\hline
	\end{tabular}
    \caption{\rewritethree{68\% completeness energies for all the stars in our 3400-3940\,K samples. This includes both the flaring the non-flaring stars. The final column indicates whether this star was used in our calculation of the flare occurrence rate in Sect.\,\ref{sec:m_star_rates}.}}\label{tab:full_early_m_stars} 
\end{table*}

\renewcommand{\arraystretch}{1.0}
\bsp	
\label{lastpage}
\end{document}